# Relative stability of a ferroelectric state in $(Na_{0.5}Bi_{0.5})TiO_3$-based compounds under substitutions: Role of a tolerance factor in expansion of the temperature interval of stable ferroelectric state

V.M. Ishchuk[1], D.V. Kuzenko[1], and V.L. Sobolev[2]

[1] *Science & Technology Center 'Reactivelectron' of the Ukrainian National Academy of Sciences, Donetsk, 83049, Ukraine*
[2] *Physics Department, South Dakota School of Mines and Technology, Rapid City, SD 57701*

The influence of the *B*-site ion substitutions in $(1-x)(Bi_{1/2}Na_{1/2})TiO_3 – xBaTiO_3$ system of solid solutions on the relative stability of the ferroelectric and antiferroelectric phases has been studied. The ions of zirconium, tin, along with $(In_{0.5}Nb_{0.5})$, $(Fe_{0.5}Nb_{0.5})$, $(Al_{0.5}V_{0.5})$ ion complexes have been used as substituting elements. An increase in the concentration of the substituting ion results in a near linear variation in the size of the crystal lattice cell. Along with the cell size variation a change in the relative stability of the ferroelectric and antiferroelectric phases takes place according to the changes of the tolerance factor of the solid solution. An increase in the tolerance factor leads to the increase in the temperature of the ferroelectric-antiferroelectric phase transition, and vice versa. All obtained results demonstrate the predominant influence of the ion size factor on the relative stability of the ferroelectric and antiferroelectric states in the $(Na_{0.5}Bi_{0.5})TiO_3$-based solid solutions and indicate the way for raising the temperature of the ferroelectric-antiferroelectric phase transition.

# 1. INTRODUCTION

The lead zirconate-titanate (PZT), lead magnesium niobate, and lead zinc niobate and solid solutions on their base are known to have the largest values of piezoelectric parameters at present. PZT-based solid solutions are the most popular commercial ceramic materials for different applications. The need to replace the lead-containing compounds has inspired an interest and stimulated intensive studies of the $(Bi_{0.5}Na_{0.5})TiO_3$ (BNT) based piezoelectric ceramics.

Majority of researchers hold to an accepted point of view that the following sequence of the structural phase transformations described in [1] takes place in BNT:

$$R3c \longleftrightarrow R3c + P4bm \longleftrightarrow P4bm \longleftrightarrow Pm3m.$$

The low-temperature rhombohedral phase ($R3c$) is a ferroelectric (FE). This state persists at heating up to ~ 200 °C. In the interval of temperatures from ~ 200 to ~ 330 °C the stable state with tetragonal crystal structure ($P4bm$) is an antiferroelectric (AFE). At higher temperatures BNT has a cubic crystal structure ($Pm3m$) and the dipole ordering does not exist. For the whole temperature range above 330 °C the dipole state of BNT is defined as a paraelectric (PE) (independent of high-temperature structural transition). There exist intermediate regions of temperature within which adjacent phases coexist in the BNT and BNT-based solid solutions.

Among BNT-based solid solutions $(1-x)(Bi_{1/2}Na_{1/2})TiO_3 – xBaTiO_3$ (BNT–100$x$BT) is the most thoroughly investigated [2-5]. Its promising piezoelectric behavior is attributed to the existence of a morphotropic phase boundary in the range of composition $x$ = 0.05 – 0.07, similar to the morphotropic boundary in the PZT [6-8]. Nonetheless these solid solutions have a number of disadvantages, such as a low temperature (of the order of 130 °C) of the disappearance of the polar FE state in solid solutions belonging to the morphotropic region and low values of the piezoelectric module $d_{33}$ (160 – 180 pC/N) .

In the majority of publications on properties of BNT-based solid solutions the samples have been obtained by substitutions in the $A$-site of the perovskite crystal lattice. But in all these cases this substitution led to a depression of the FE – AFE phase transition temperature (in comparison with the value of this temperature in the prime BNT). The variation of the solid solution's depolarization temperature with changes in the content of divalent and trivalent elements used for substitutions in the A-site is given in Fig.1 from [9]. Moreover, even in the BNT-based solid solutions containing the high-temperature ferroelectrics such as PbTiO$_3$ or BiFeO$_3$, the point of destruction of the polar FE state decreased (as in $(Bi_{1/2}Na_{1/2})TiO_3$ – BiFeO$_3$) solid solution [10]), or just did not increase (as in $(Bi_{1/2}Na_{1/2})TiO_3$ – PbTiO$_3$ solid solution [11, 12]).

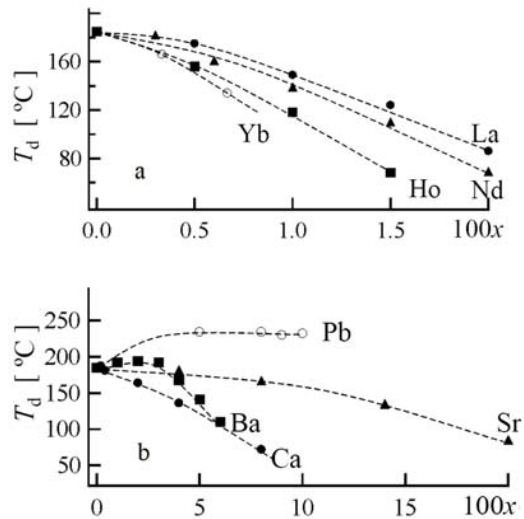

Fig.1. Depolarization temperature, $T_d$, as a function of the content of substituting ions, $x$, in: $[(Bi_{1-x}Ln^{3+}_x)_{1/2}Na_{1/2}]TiO_3$ (a) and $[(Bi_{1/2}Na_{1/2})_{1-x}A^{2+}_x]TiO_3$ (b) solid solutions [9].

Large values of the piezoelectric parameters in lead-containing piezoelectric substances are reached as a consequence of *B*-site substitutions. PZT, $Pb(Mg_{1/3}Nb_{2/3})O_3 – PbTiO_3$, and $Pb(Zn_{1/3}Nb_{2/3})O_3 – PbTiO_3$ solid solutions are good examples of this statement. In PZT solid solutions one of the components ($PbTiO_3$) is a ferroelectric and the other one ($PbZrO_3$) is an antiferroelectric. An increase in the concentration of a substituting ion that has a larger size than the size of the titanium ion leads to the stabilization of a non-piezoelectric AFE phase [13-16]. Conversely a decrease in the size of the ion substituting for the *B*-site ion leads to the stabilization of a FE state.

There are very few reports in the literature (see, for example, [17-19]) on the BNT-based solid solutions with substitutions of zirconium for titanium. The number of publications devoted to *B*-site substitutions by other ions is even less [20-23]. As expected the AFE state in BNT-BT was also stabilized [24, 25] when the substitutions of ions with a size larger than the size of the $T^{+4}$-ion took place. As mentioned above substitutions of smaller ions in *B*-site of crystal lattice can stabilize the FE state. Therefore, for investigations of the influence of substitutions for titanium ion on the relative stability of the FE and AFE states we have chosen the $[(Na_{0.5}Bi_{0.5})_{0.8}Ba_{0.2}]TiO_3$ as a base solid solution. The solid solution with such composition is located within the FE region in the diagram of the phase states of the $[(Na_{0.5}Bi_{0.5})]TiO_3 − BaTiO_3$ system [2, 26]. This compound undergoes a phase transition from the FE state into the AFE state when heated and has a subsequent phase transition into the PE state at higher temperatures. We have also studied solid solutions with smaller barium content located in the vicinity of the morphotropic region.

The aim of our studies is to investigate the influence of the size of ions substituting the *B*-site in the chosen system of BNT-based solid solutions on the relative stability of the FE and AFE states and to determine the physical factors that promote the increase of the temperature range of existence of the FE state.

In this paper we present studies of the solid solutions with compositions located in two different areas of the diagram of phase states of $[(Na_{0.5}Bi_{0.5})]TiO_3–BaTiO_3$, namely, solid solutions with compositions positioned away from the morphotropic region (morphotropic region of composition is near the 20 mol% of $BaTiO_3$) and solid solutions with composition in the vicinity the morphotropic region.

## 2. EXPERIMENTAL METHODS

The traditional solid state synthesis method has been used for manufacturing of BNT-based solid solutions. Reagent grade oxides and carbonates of corresponding metals were used as starting materials that were mixed in the appropriate stoichiometry (except for $Bi_2O_3$, which we have taken in the excess of 0.5 wt %) by ball milling during 12 h. The mixed powders were calcined at 850 $^oC$ for 6 h. The sintering of ceramic samples was carried out at 1200 $^oC$ for 6 h.

The studies of crystal structure were performed using ДРОН-3 X-ray diffractometer in the Bragg-Brentano geometry with $CuK_\alpha$-radiation ($\lambda$ = 1.5418 Å), Ni-filter for the incident beam and the graphite monochromator in the diffracted beam. The angular range was $20^o \leq 2\theta \leq 90^o$, the scan step was $0.02^o$, and the accumulation time in each point was 2 s. The X-ray diffraction patterns showed perfect single-phase state of the samples prepared for further studies. X-ray diffraction patterns for the solid solutions with the zirconium substituted for titanium are presented in Fig.2 as an example.

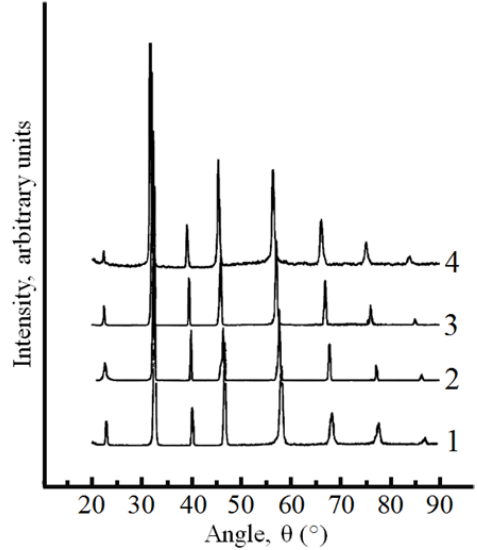

Fig.2. X-ray diffraction patterns for the $[(Na_{0.5}Bi_{0.5})_{0.80}Ba_{0.20}](Ti_{1-y}Zr_y)O_3$ system of solid solutions. Zr-content ($y$): 1 – 0.00, 2 – 0.05, 3 – 0.10, 4 – 0.30.

We have given here this example, since the process of obtaining of single-phase solid solutions when zirconium is substituted for titanium is quite complicated [24, 25]. This is due to the circumstance that the $Ti^{4+}$ and $Zr^{4+}$ ions have the largest difference in sizes in comparison with all other substituting ions studied in the present paper.

Synthesized powders were axially pressed into disks with a diameter of 12 mm and a thickness of 1 mm and then sintered at 1150 – 1200 $^{\circ}C$ to manufacture samples for dielectric measurements. Fire-on silver paste was used to make electrodes on disk's surfaces. Temperature dependencies of the dielectric constant, $\varepsilon(T)$, were measured in an AC electric field with a frequency of 1 kHz using a precision QuadTech 7600 LCR meter. D-E hysteresis loops were observed using the standard Sawyer-Tower circuit at $10^{-2}$ Hz.

## 3. INFLUENCE OF ION SUBSTITUTIONS ON PHASE TRANSFORMATIONS

### 3.1. $[(Na_{0.5}Bi_{0.5})_{0.8}Ba_{0.2}]TiO_3$ based solid solutions

As noted before our goal is to study the influence of substitutions on the relative stability of the FE and AFE states in BNT-100xBT solid solutions located in the vicinity of the morphotropic (MPB) region of the BNT–100xBT phase diagram. We also consider the solid solutions containing 20 mol% of $BaTiO_3$ since in these compounds the FE state is more stable than the AFE one. Here, we also need to note that as a rule a little attention has been paid to the region of AFE states in publications on BNT–BT compounds. Such situation could be due to the desire of finding the substitute for PZT-based materials and majority of studies were devoted to the MPB region of the BNT–100xBT phase diagram. Because of this we give a discussion of BNT–BT solid solutions with large barium content.

The first "composition−temperature" diagram of phase states for the $(1-x)(Bi_{1/2}Na_{1/2})TiO_3$ – $xBaTiO_3$ system of solid solution was presumably published in [26]. This phase diagram was rather schematic reflecting only general tendencies in the formation of different phase states in this series of solid solutions. The region of the AFE states located between the regions of the FE and paraelectric states was shown in this phase diagram. The study [2] contained the most complete phase diagram for ceramic samples of solid solutions with BaTiO3-content in the interval from 0.00% to 0.15%. Authors of this paper demonstrated that the temperature region for the AFE states is located between the temperature regions for the FE and paraelectric states. There were also additional phase boundaries present within the region of the AFE states in this phase diagram [2]. Probably their presence was a feature of the AFE region itself but it could also be a consequence of using of ceramic samples for all studies. The phase diagram obtained on single crystals of the same series of $(Bi_{1/2}Na_{1/2})1-xBaxTiO_3$ system of solid solutions was published in [27]. High precision investigations of crystal structure using synchrotron radiation as well as dielectric measurements on crystals in a wide temperature interval were carried out in this study. Authors of [27] have also indicated the presence of the intermediate region of AFE states. It must be emphasized that the authors of [2] did not find any additional transitions within the region of AFE states or they did not mention their presence. Summarizing all this, one can conclude that that the vast majority of authors have clearly pointed out the existence of the AFE states in the BNT–100xBT solid solutions with $BaTiO_3$-content up to 15%.

Very few publications on BNT–100xBT solid solutions with higher content of barium titanate are available. As we mentioned above the reason for this may be due to the primary interest in BNT–100xBT solid solutions from the MPB region.

The boundaries of the AFE states region do not collapse when the concentration of $BaTiO_3$ reaches 15% and continue to be present even at higher $BaTiO_3$-content (this AFE region is an intermediate area between the FE and paraelectric regions in the phase diagram). There are very few publication devoted to studies of the BNT–100xBT compounds with higher content of $BaTiO_3$. The whole series of BNT–100xBT solid solutions with $BaTiO_3$-content up to 90% was investigated in [28]. It was shown that there is a phase transition in the temperature region of 160°C–170°C besides of the phase transition into the paraelectric state at 300°C in the solid solution with 20% of $BaTiO_3$. Turning back to the phase diagram in [2] one can see that this transition temperature "rests" on the extension of the line of the FE-AFE phase transition. In addition to all above-said one has to mention [29] where the phase diagram of BNT–100xBT with 20% of $BaTiO_3$ is presented. Authors of this paper observed the presence of two phase transitions (from the FE to AFE state and from the AFE to PE state at heating) in this solid solution. If to compare this phase diagram with all other known phase diagrams one can clearly see that this intermediate state between the FE and paraelectric states is the AFE state.

Our discussions of the AFE state and the FE-AFE phase transition in this study has not been based on the above-mentioned results only. We carried out measurements of temperature dependencies of the dielectric characteristics of our samples subjected to the action of an external electric field. Temperature dependencies of the imaginary part of dielectric constant measured without external electric field and in the presence of a field of 1100 V/mm are given in Fig. 3. Here we need to note that the anomaly in the $\varepsilon^{//}(T)$ dependence in the vicinity of the paraelectric phase transition is very weakly pronounced and is not visible on a scale of Fig. 3.

The temperature of the low-temperature transition increases with an increase in the electric field intensity at a rate of 19°C/(kV/mm) while the temperature of the high-temperature transition decreases at a rate of 1.5°C/(kV/mm). Such type of changes in the temperatures of the phase transitions in the presence of external electric field and a weak decrease in the temperature of paraelectric transition are characteristic specifically for the state with the AFE-type of electric-dipole ordering [16, 30, 31].

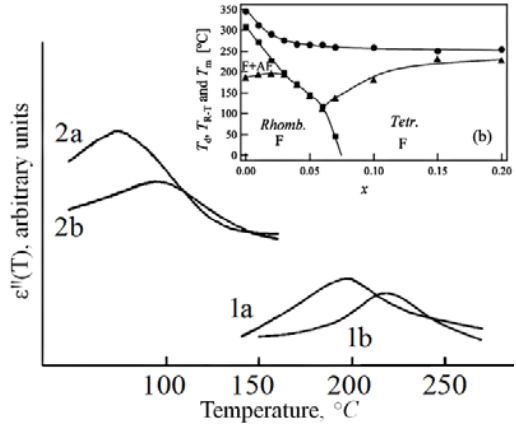

Fig.3. Imaginary part of the dielectric constant vs. temperature for the $(Na_{0.5}Bi_{0.5})_{0.80}Ba_{0.20}](Ti_{1-y}Zr_y)O_3$ solid solutions (a) without and (b) under the action of the an electric field of 1100 V/mm.
Zr-content ($y$): 1 – 0.00, 2 – 0.05.
The "composition-temperature" phase diagram for the $[(Na_{0.5}Bi_{0.5})_{1-x}Ba_x]TiO_3$ series from [29] is shown in the insert.

The rate of displacement of the temperature of the low-temperature phase transition (which we identify as the FE–AFE phase transition) in external electric field observed in our measurements is close to the rate of displacement of the FE–AFE phase transition in solid solutions with small BaTiO$_3$-content where the AFE phase has been unambiguously identified. These results are presented in Figure 2 of [32]. The influence of the Zr-ion content on the position of the high-temperature AFE–PE phase transition is weak. The temperature, at which the low-temperature FE–AFE phase transition takes place, decreases significantly when the content of zirconium ion in the solid solution increases.

At the first stage, we have carried out substitutions of zirconium and tin ions for titanium ions. The results of such substitutions have been discussed in papers [24, 25] in details.

In the $[(Na_{0.5}Bi_{0.5})_{0.80}Ba_{0.20}](Ti_{1-y}Zr_y)O_3$ (the Zr ionic radius is larger than the one of titanium) solid solutions with different zirconium content the two phase transitions were clearly observed. The first of them (the low temperature one) corresponds to the FE-AFE phase transition and the second one corresponds to the transition from the AFE (non-polar) to the paraelectric phase (the attribution of phases was made according to the results of [2, 28, 29, 33]). The temperature, at which the low temperature phase transition takes place, decreases significantly when the zirconium content in the solid solution rises. The effect of variation of the zirconium concentration on the position of the high temperature phase transition is weak.

The study of the crystal structure of solid solutions showed that an increase in the zirconium content led to a linear increase in the size of the unit cell and to a rapid decrease in the degree of tetragonal distortions of the crystal cell, and at x > 0.10 it was difficult to say anything about the type of distortion of the crystal lattice. Fig.4 shows the dependencies of the crystal lattice parameter, calculated in pseudocubic approximation, on the zirconium content. As we can see, the linear dependence is performed with good accuracy.

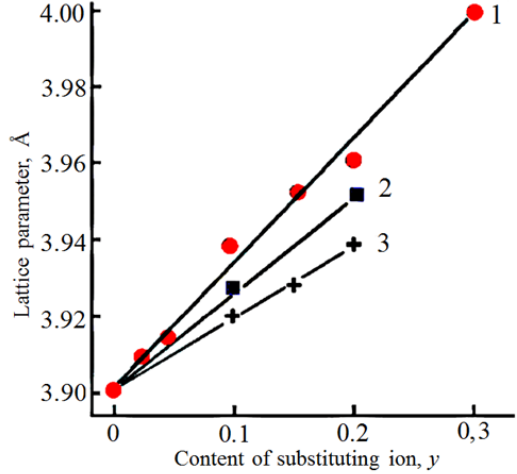

Fig.4. The dependence of the crystal lattice parameter (in pseudocubic approximation) on the content of substituting element, $y$, in the $(Na_{0.5}Bi_{0.5})_{0.80}Ba_{0.20}](Ti_{1-y}B_y)O_3$.
Subsisting elements: 1 − Zr, 2 − Sn, 3 − $(Fe_{1/2}Nb_{1/2})$.

Similar results were obtained for solid solution with substitutions of titanium by tin which has an ionic radius smaller than the one of zirconium on but still larger than the ionic radius of titanium. The growth of the tin content led to an increase in the unit cell size (Fig.4), however, since the size of the tin ion is smaller than that of zirconium, the rate of the lattice parameter increase was lower in this case. The temperature of the phase transition between the FE and AFE states decreased when the tin content increased [24].

The influence of zirconium and tin ions substitutions on the relative stability of the FE and AFE states can be explained on the basis of the size effect, as it has been done in relation to the stability of the same phases in the PZT-based solid solutions. The tolerance factor $t = (R_A + R_O)/\sqrt{2}(R_B + R_O)$ is usually taken into account as an estimation parameter. The FE state in the PZT is stable at large values of $t$ ($t > 0.9090$), whereas for small values of $t$ ($t < 0.9080$) the stable state is the AFE one [34]. As it easy to see, an increase in the ionic radius of the $B$-site ion (when ions $Ti^{4+}$ are substituted by $Zr^{4+}$, $Sn^{4+}$ ions) leads to a decrease of the $t$-factor. Just this effect occurs in the case of substitutions of zirconium (as well as by tin) for titanium in $(Na_{0.5}Bi_{0.5})TiO_3$ based solid solutions. Such a conclusion has been confirmed by results obtained on the BNT–100xBT solid solutions with substitutions of the $Ti^{4+}$ ion by the complex $(In_{1/2}Nb_{1/2})^{4+}$. The average ionic radius of the $(In_{1/2}Nb_{1/2})^{4+}$ complex is larger than the radius of $Ti^{4+}$ ion. According to approach based on consideration of the tolerance factor, the stability of the FE state has to be depressed against the stability of the AFE state, when the content of $(In_{1/2}Nb_{1/2})^{4+}$ complex rises, and the temperature of FE-AFE phase transition decreases. Such behavior has been observed in our experiments.

### 3.2 Substitutions of the $(Fe_{0.5}Nb_{0.5})$ complex for titanium

As the next step of our studies to verify the proposed mechanism of influence of the size of ionic radius of substituting ions on the FE→AFE phase transition in the BNT-based solid solutions we carried out substitutions of both zirconium and titanium ions by the $(Fe_{0.5}Nb_{0.5})^{4+}$ complex with the same valence as the replaceable ions. Two samples of the BNT-100xBT system were obtained $[(Na_{0.5}Bi_{0.5})_{0.80}Ba_{0.20}]\{[(Ti_{0.90}(Fe_{0.5}Nb_{0.5})_{0.1})]_{0.90}Zr_{0.10}\}O_3$ (the substitution of the $(Fe_{0.5}Nb_{0.5})^{4+}$ complex for $Ti^{4+}$) and $[(Na_{0.5}Bi_{0.5})_{0.80}Ba_{0.20}]\{(Ti_{0.90}[Zr_{0.08}(Fe_{0.5}Nb_{0.5})_{0.02}]_{0.10}\}O_3$ (the substitution of the $(Fe_{0.5}Nb_{0.5})^{4+}$ complex for $Zr^{4+}$). In the first case, the average ionic radius of the $(Fe_{0.5}Nb_{0.5})^{4+}$ complex is greater than the radius of the substituted titanium ion; in the second case the average ionic radius of $(Fe_{0.5}Nb_{0.5})^{4+}$ is smaller than the radius of zirconium ion. Fig.5 shows the temperature dependencies of the imaginary part of the dielectric constant for these solid solutions. As one can see, in the first case (in full agreement with the proposed model) the stability of the AFE state with respect to the FE state increases, and the FE→AFE transition temperature decreases. In the second case both the stability of the FE state and the temperature of the FE→AFE phase transition also increases.

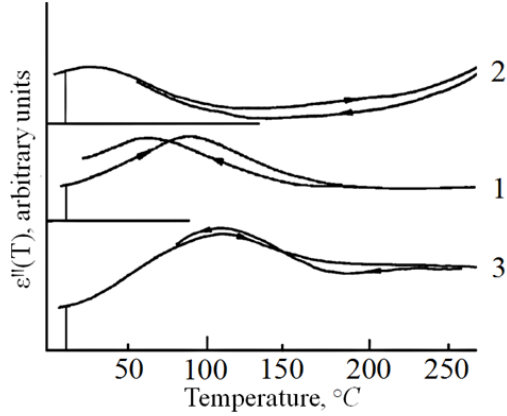

Fig.5. Imaginary part of the dielectric constant vs. temperature for the $(Na_{0.5}Bi_{0.5})_{0.80}Ba_{020}](Ti_{0.9}Zr_{0.10})O_3$ based solid solutions.
1 – dependence for the $(Na_{0.5}Bi_{0.5})_{0.80}Ba_{020}](Ti_{0.9}Zr_{0.10})O_3$ without substitutions for Ti- and Zr
2 – dependence for the $(Na_{1/2}Bi_{1/2})_{0.80}Ba_{0.210}][\{[Ti_{0.9}(Fe_{0.5}Nb_{0.5})_{0.10}]_{0.90}Zr_{0.10}\}O_3$ with substitution of $(Fe_{0.5}Nb_{0.5})$ complex for Ti
3 – dependence for the $[(Na_{1/2}Bi_{1/2})_{0.80}Ba_{0.20}]\{Ti_{0.90}[Zr_{0.08}(Fe_{0.5}Nb_{0.5})_{0.02}]_{0.10}\}O_3$ with substitution of $(Fe_{0.5}Nb_{0.5})$ complex for Zr.

### 3.3. $[(Na_{0.5}Bi_{0.5})]TiO_3 − BaTiO_3$ with compositions in the vicinity of the morphotropic region

Results presented in previous section demonstrate the possibility for increasing the stability of the FE state relative to the AFE one and as consequence a possibility to increase the temperature of the FE-AFE phase transition. To achieve this one has to substitute the titanium ion by ions or ion complexes with smaller ionic radii.

The search for ions of chemical elements or ion complexes that can stabilize the FE state with the only titanium substitution was the next stage of our research. There is no suitable ions of simple chemical elements which have size lower than the size of $Ti^{+4}$ and are specific for B-site of the perovskite crystal lattice of the $[(Na_{0.5}Bi_{0.5})_{1-x}Ba_x]TiO_3$ solid solutions. Therefore we have selected the ion complex $\left(Al^{3+}_{0.5}V^{5+}_{0.5}\right)^{4+}$ with an average ion radius of 0.54 Å which is smaller then the radius of the $Ti^{4+}$ ion equal to 0.605 Å. We have analyzed two base solid solutions in our studies, namely, the $[(Na_{0.5}Bi_{0.5})_{0.97}Ba_{0.03}]TiO_3$ and $[(Na_{0.5}Bi_{0.5})_{0.90}Ba_{0.10}]TiO_3$ (with 3 and 10 % of Ba-content). The compositions of these compounds are located in the vicinity the morphotropic region in phase diagram of BNT-100xBT. The first of these solid solutions is located in the region of the phase diagram containing solid solutions with rhombohedral crystal structure. The second compound belongs to the region of solid solutions with tetragonal crystal structure.

The replacement of $Ti^{4+}$- ions by the complex $\left(Al^{3+}_{0.5}V^{5+}_{0.5}\right)^{4+}$ does not lead to a change in the type of distortion of the perovskite crystal lattice. So the solid solution $[(Na_{0.5}Bi_{0.5})_{0.97}Ba_{0.03}][Ti_{1-x}(Al_{0.5}V_{0.5})_x]O_3$ with $x = 0.10$ is characterized by the following parameters of the rhombohedral unit cell: $a_{Rh} = 3.887$ Å and $\alpha_{Rh}$ = 89.75°. The $[(Na_{0.5}Bi_{0.5})_{0.90}Ba_{0.10}][Ti_{1-x}(Al_{0.5}V_{0.5})_x]O_3$ solid solution with $x$ = 0.10 is characterized by parameters of the tetragonal unit cell: $a_T$ = 3.905 Å, $c_T$ = 3.952 Å and $c/a$ = 1.012. The degree of distortion of the crystal lattice is somewhat higher (albeit insignificantly) than in unsubstituted solid solutions [35]. It will be demonstrated below that the introduction of the $\left(Al_{0.5}^{3+}V_{0.5}^{5+}\right)^{4+}$ ion complex into the crystal lattice leads to an increase in the temperature of destruction of the FE ordering. Thus, substitutions of the $\left(Al_{0.5}^{3+}V_{0.5}^{5+}\right)^{4+}$ ion complex increase of the stability of the FE phase against the AFE one. The latter define the increase in the degree of distortion of the crystal lattice.

An example of temperature dependencies of the dielectric constant for the series of solid solutions with 10% of barium in the $A$-sites of the perovskite crystal lattice of $[(Na_{0.5}Bi_{0.5})_{0.90}Ba_{0.10}][Ti_{1-x}(Al_{0.5}V_{0.5})_x]O_3$ are shown in Fig.6. As it is clearly seen, an increase of the $\left(Al_{0.5}^{3+}V_{0.5}^{5+}\right)^{4+}$ content leads to the rise of the temperature of the FE-AFE phase transition. The influence the substitution of the $\left(Al_{0.5}^{3+}V_{0.5}^{5+}\right)^{4+}$ complex for $Ti^{4+}$ on the phase transition between the FE and AFE phases in the solid solutions containing 3% of $Ba^{2+}$ has an analogous character.

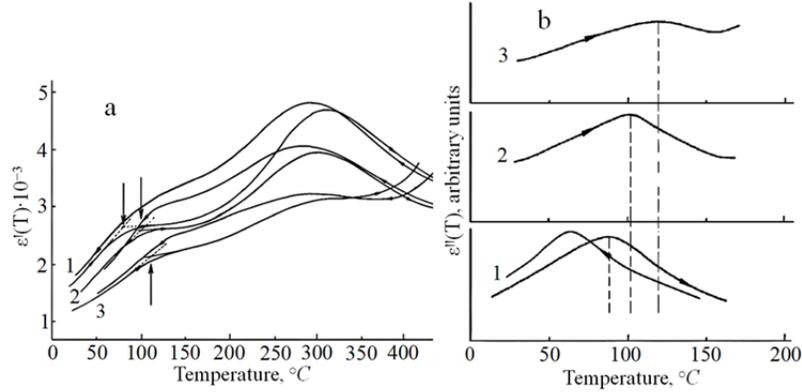

Fig.6. Real (*a*) and imaginary (*b*) parts of the permittivity vs. temperature for the $(Na_{0.5}Bi_{0.5})_{0.90}Ba_{0.10}][Ti_{1-y}(Al_{0.5}V_{0.5})_y]O_3$ solid solutions.
*y*: 1 – 0.00, 2 – 0.05, 3 – 0.10.
The arrows in Fig. 6a indicate the temperatures of the FE-AFE transition.

The dependence of the temperature of the FE-AFE phase transition on the content of $\left(Al_{0.5}^{3+}V_{0.5}^{5+}\right)^{4+}$ for the two series of the solid solutions is shown in Fig.7. The larger rate of increase of the FE-AFE transition temperature has been observed for the series of solid solutions containing 3% of the barium.

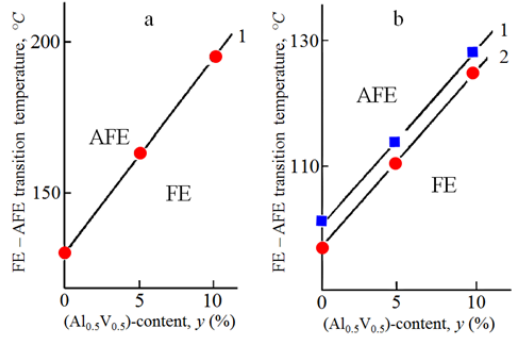

Fig.7. The FE-AFE phase transition temperature vs content of the $(Al_{0.5}V_{0.5})$ complex:
A – in the $[(Na_{0.5}Bi_{0.5})_{0.0.97}Ba_{0.0.03}][Ti_{1-y}(Al_{0.5}V_{0.5})_y]O_3$ solid solutions
b – in the $[(Na_{0.5}Bi_{0.5})_{0.0.90}Ba_{0.0.10}][Ti_{1-y}(Al_{0.5}V_{0.5})_y]O_3$ solid solutions during the heating (1) and cooling (2).

## IV. CONCLUSIONS

We have investigated the influence of ion substitutions in the *B*-site of the crystal lattice of $(Na_{0.5}Bi_{0.5})TiO_3$-based solid solutions on the temperature of the FE − AFE phase transition.

An increase in the content of the substituting ions results in the variation of the size of the crystal cell according to a nearly linear law. At the same time there exists a coordination of the relative stability of the AFE and FE phases with the variation of the tolerance factor of the solid solution.

The substitution of titanium ion by ions of zirconium and tin or by $(In_{0.5}Nb_{0.5})$ complex leads to a decrease in the tolerance factor and to an increase of the stability of the AFE phase relative to the stability of the FE phase.

The influence of substitutions of the $(Fe_{0.5}Nb_{0.5})$ complex depends on the type of the ion in the *B*-cite which is substituted by this complex. When the $(Fe_{0.5}Nb_{0.5})$ complex is substituted for a zirconium ion the tolerance factor increases and the temperature of the phase transition rises (it means the increase of the FE phase stability), whereas when the $(Fe_{0.5}Nb_{0.5})$ complex is substituted for a titanium ion the transition temperature decreases as the result of a decrease of the tolerance factor (it means the increase of the AFE phase stability).

The substitution of the ion in the *B*-cite by the $(Al_{0.5}V_{0.5})^{4+}$ complex leads to the increase of the stability of the FE phase and the increase in the temperature of the FE−AFE phase transition.

All results presented in this paper show a predominant influence of the ion size factor on the relative stability of the FE and AFE states in the $(Na_{0.5}Bi_{0.5})TiO_3$-based solid solutions in the case of *B*-site substitutions and point at the method allowing to raise the temperature of the FE−AFE phase transition.

References

[1] G. O. Jones and P. A. Thomas, Acta Cryst. B **58**, 168 (2002).
[2] F. Cordero, F. Craciun, F. Trequattrini, E. Mercadelli and C. Galassi, Phys. Rev. B **81**, 144124 (2010).
[3] Y.-Q. Lu and Y.-X. Li, J. Adv. Dielectr. **1**, 269 (2011).
[4] H. D. Li, C. D. Feng, and W. L. Yao, Mater. Lett. **58**, 1194 (2004).
[5] E. Sapper, N. A. Novak, W. Jo, T. Granzow, and J. Rödel, J. Appl. Phys. **115**, 194104 (2014).
[6] P. K. Panda, J. Mater. Sci. **44**, 5049 (2009).
[7] J. Rödel, W. Jo, K. T. P. Seifert, E. M. Anton, T. Granzow, and D. Damjanovic, J Amer. Ceram. Soc. **92**, 1151 (2009).
[8] B. Parija, S. Panigrahi, T. Badaranda, and T.P. Sinha. J. Adv. Dielectrics **2**, 125008 (2012).
[9] Y. Hiruma, Y. Watanabe, H. Nagata, and T. Takenaka, Key Engineering Materials **350**, 93 (2007).

[10] J. M. Kim , Y. S. Sung , J. H. Cho , T. K. Song , M. H. Kim , H. H. Chong , T. G. Park , D. Do, and S. S. Kim, Ferroelectrics **404**, 88 (2010).
[11] S.-E. Park and K.S.Hong, J. Appl. Phys. **79**, 383 (1996).
[12] K.S.Hong and S.-E. Park, J. Appl. Phys. **79**, 388 (1996).
[13] B. Jaffe, W.R. Cook, and H. Jaffe, *Piezoelectric Ceramics* (Academic Press, London and New York, 1971) Ch. 7, 8.
[14] M. Troccaz, P. Gonnard, Y. Fettineau, and L. Eyraud, Ferroelectrics **14**, 769 (1976).
[15] P. Gonnard and M. Troccaz, J. Solid State Chem. **23**, 321 (1978).
[16] M. E. Lines and A.M. Glass, *Principles and Application of Ferroelectric and Related Materials* (Clarendon Press, Oxford, 1977) Ch. 8.
[17] L.K. Kumari, K. Prasad, and R.N.P. Choudhary, Indian J. Energ. Mater.Sci. **15**, 147 (2008).
[18] C. Peng, J.-F. Li, and W. Gong, Mater. Lett. **59**, 1576 (2005).
[19] B. Tilak, Amer. J. Mater. Sci. **2**, 110 (2014).
[20] E. Aksel, J.S. Forrester, B. Kowalski, and M.Deluca, Phys. Rev. B **85**, 024121 (2012).
[21] W. Bai, Y. Bian, J. Hao, B. Shen, and J. Zhai, J. Am. Ceram. Soc. **96**, 246 (2013)
[22] E.A. Patterson and D.P. Cann, J. Am. Ceram. Soc. **95**, 3509 (2012)
[23] J. H. Cho, S. C. Lee, L. Wang, H. G. Yeo, Y. S. Sung, M. H. Kim, and T. K. Song, J. Korean Phys. Soc. **56**, 457 (2010).
[24] V.M. Ishchuk, L.G. Gusakova, N.G. Kisel, D.V. Kuzenko, N.A. Spiridonov, and V.L. Sobolev, Mat. Res. Express **3**, 2 (2016).
[25] V.M. Ishchuk, L.G. Gusakova, N.G.Kisel, N.A. Spiridonov, and V.L. Sobolev, J. Amer. Ceram. Soc. **99**, 1786 (2016).
[26] T. Takenaka, K. Maruyama, and K. Sakata, Jpn. J. Appl. Phys., Part 1 **30**, 2236 (1991).
[27] W. Ge, C. Luo, Q. Zhang, Y. Ren, J. Li, H. Luo, and D. Viehland, Appl. Pys. Lett. **105**, 162913 (2014).
[28] M. Dunce , E. Birks , M. Antonova , A. Plaude , R. Ignatans, and A. Sternberg, Ferroelectrics **447**, 1 (2013)
[29] Y. Hiruma, K. Yoshii, H. Nagata, and T. Takenaka, Ferroelectrics **346**,114 (2007)
[30] G. Shirane, E. Savaguchi, and Y. Tagaki, Phys. Rev. **84**, 476 (1951).
[31] J.C. Burfoot and G.W. Taylor, *Polar Dielectrics and Their Applications* (University of California Press, Berkeley, 1979).
[32] E. Sapper, S. Schaab, W. Jo, T. Granzow, and J. Röodel, J. Appl. Phys. **111**, 014105 (2012).
[33] R. Garg, B. N. Rao, A. Senyshyn, P. S. R. Krishna, and R. Ranjan, Phys. Rev. B **88**, 014103 (2013).
[34] M. Troccaz, P. Gonnard, Y. Feteveau, and L. Eyraud, Ferroelectrics, **14**, 769 (1976).
[35] W. Jo, J.E. Daniels, J.L. Jones, X. Tan, P.A. Thomas, D. Damjanovic, and J. Rödel, J. Appl. Phys. **109**, 014110 (2011).